\documentclass[twocolumn,showpacs,preprintnumbers] {revtex4-2}
\usepackage{graphicx,latexsym,amssymb,amsmath,color,multirow,mathrsfs,booktabs,ifpdf,epsfig,dcolumn,bm,longtable}
\usepackage{changes}
\usepackage{hyperref}
\hypersetup{colorlinks=true,linkcolor=blue,filecolor=magenta,urlcolor=cyan}
\date{\today}

\usepackage{graphicx}
\usepackage{dcolumn}
\usepackage{bm}

\date{\today}

\begin{document}



\title{Single-particle properties of the near-threshold proton-emitting resonance in $^{11}$B}

\author{Nguyen Le Anh}
\email{anhnl@hcmue.edu.vn}
\affiliation{Department of Theoretical Physics, Faculty of Physics and Engineering Physics, University of Science, Ho Chi Minh City, Vietnam}
\affiliation{Vietnam National University, Ho Chi Minh City, Vietnam}
\affiliation{Department of Physics, Ho Chi Minh City University of Education, 280 An Duong Vuong, District 5, Ho Chi Minh City, Vietnam}
\author{Bui Minh Loc}%
\email{buiminhloc@ibs.re.kr}
\affiliation{Center for Exotic Nuclear Studies, Institute for Basic Science (IBS), Daejeon 34126, Korea}
\author{Naftali Auerbach}
\email{auerbach@tauex.tau.ac.il}
\affiliation{School of Physics and Astronomy, Tel Aviv University, Tel Aviv 69978, Israel}%
\author{Vladimir Zelevinsky}
\email{Zelevins@frib.msu.edu}
\affiliation{Facility for Rare Isotope Beams, Michigan State University, East Lansing, Michigan 48824, USA}
\affiliation{National Superconducting Cyclotron Laboratory, Michigan State University, East Lansing, Michigan 48824, USA}
\affiliation{Department of Physics and Astronomy, Michigan State University, East Lansing, Michigan 48824, USA}

\begin{abstract}
The excitation function of proton elastic scattering from $^{10}$Be at keV energy is calculated using the self-consistent Skyrme Hartree-Fock in the continuum method. The calculation successfully reproduces the narrow near-threshold proton-emitting resonance ($E_x = 11.4$ MeV, $\Gamma = 6$ keV, and quantum number $J^{\pi} = 1/2^+$) in $^{11}$B relevant to the $\beta$-delayed proton emission of $^{11}$Be. This supports the recent experimental result of Y. Ayyad \textit{et al.} at the ReA3 re-accelerator facility of the National Superconducting Cyclotron Laboratory (NSCL) at the Michigan State University. The resonance is interpreted as the $s_{1/2}$ single-proton resonance state in the Skyrme Hartree-Fock mean-field theory.
\end{abstract}

\maketitle

\textit{Introduction.} The study of loosely bound states of atomic nuclei close to the drip lines and beyond is one of the major current directions of nuclear science \cite{Johnson2020}.
At limits of nuclear stability, rare decay processes were discovered for weakly bound nuclei, especially for nuclei with the halo structure (see \cite{Pfutzner2012} for a review).

A very particular case is the $\beta$-delayed proton emission ($\beta p$) that was theoretically predicted for the halo nucleus $^{11}$Be in Refs.~\cite{Horoi2003, Baye2011} with the unexpectedly high intensity \cite{Baye2011}. The decay was indirectly observed in Refs.~\cite{Riisager2014, Borge2013, Riisager2020}. The existence of a narrow resonance in $^{11}$B was suggested to explain this observation \cite{Riisager2014,Riisager2014NPA}. However, there was no suitable level in $^{11}$B known at that time. There have been several attempts of theoretical calculations to confirm the resonance in $^{11}$B and to estimate the $\beta^-p$ decay branching ratio in $^{11}$Be \cite{Volya2020,Okolowicz2020,Okolowicz2021,Elkamhawy2021}. These studies are also related to the search for the so-called neutron dark decay mode proposed in Ref.~\cite{Fornal2018} in order to explain the existing discrepancy between two different methods of neutron lifetime measurements (see \cite{Wietfeldt2011} for the review). If the dark neutron decay is possible for the unbound nucleon, then it should occur also for the quasi-bound neutron in a nucleus with sufficiently low neutron binding energy, and $^{11}$Be is the most promising nucleus for such studies \cite{Pfutzner2018}.

Recently, the $\beta^- p$ decay was directly observed for the first time in the ${}^{11}$Be $\rightarrow {}^{10}$Be + $\beta^{-} + p$ \cite{Ayyad2019}. The result showed that the $\beta^{-}p$ decay is the sequence of the disintegration proceeded via an intermediate near-threshold narrow resonance in the $\beta^{-}$ decay product $^{11}$B$^{*}$ at an energy $E_r = 196(20)$ keV above the proton separation energy, with a total width of $\Gamma_p = 12(5)$ keV, and spin-parity quantum numbers $J^\pi = (1/2^{+}, 3/2^{+})$.
Moreover, the dedicated experiment employing the $^{10}$Be($p,p$) reaction was performed later by the same collaboration at the ReA3 re-accelerator NSCL facility \cite{Ayyad2022}. The experimental result indicated a very narrow resonance ($\Gamma_p = 4.4$ keV) with $J^\pi = 1/2^+$ located by 182 keV above the proton separation energy.

Using an appropriate optical potential for the keV-energy scattering, it is straightforward to calculate the excitation function of the elastic scattering $^{10}$Be($p,p$). 
The detailed calculation for the nuclear structure can be directly linked to the problem of scattering without introducing additional approximations, especially in the very low-energy region where the coupling to the inelastic levels can be neglected. In the pioneering work Ref.~\cite{Vautherin68}, the Hartree-Fock (HF) single-particle potential was expressed as an equivalent optical potential. The version of the optical potential obtained from the HF calculations using the Skyrme effective interaction was demonstrated to be an appropriate formalism for the keV-energy region \cite{Dover1971,Dover1972}. Not only the background cross section but also the positions and widths of single-particle resonances could be predicted. Recently, the method was successfully applied to the study of keV-nucleon radiative capture reactions in nuclear astrophysics \cite{Anh2021PRC103,Anh2021PRC104}.
In this work, the Skyrme HF in the continuum method is applied to reproduce the near-threshold resonance in $^{11}$B recently found in Ref.~\cite{Ayyad2022}.

\textit{Method.} The solution starts with the Skyrme HF calculation for the ground state of $^{10}$Be using the \texttt{skyrme\_rpa} computer program provided in Ref.~\cite{Colo2013}. The versions of Skyrme interactions used in this study include SkM$^*$ \cite{Bartel1982}, SGII \cite{Giai1981}, SLy4 \cite{Chabanat1998}, and SAMi \cite{Roca-Maza2012}. The local equivalent optical potential within the Skyrme HF formalism (called the Skyrme HF optical potential) $ V(E,r)$, is obtained following the method of Ref.~\cite{Dover1971}:
\begin{align}\label{Vopt}
    V(E,r) = \dfrac{m^*(r)}{m} \left\{ V_{\rm HF}(r) + \dfrac{1}{2}\dfrac{d^2}{dr^2}\left(\dfrac{\hbar^2}{2m^*(r)}\right) \right. \nonumber \\
    \left. -\dfrac{m^*(r)}{2\hbar^2} \left[\dfrac{d}{dr}\left(\dfrac{\hbar^2}{2m^*(r)}\right)\right]^2\right\} + \left[1-\dfrac{m^*(r)}{m} \right]E,
\end{align}
where $m$ is the nucleon mass, $V_{\rm HF}(r)$ and $m^*(r)$ are the mean-field potential and the effective proton mass within the Skyrme HF approach, respectively \cite{Vautherin1972}.
At the keV energy, the nuclear central potential $V(E, r)$ is real and depends weakly on the incident energy. The center-of-mass correction, the rearrangement term, and the non-local effect are included in the Skyrme HF optical potential \cite{Dover1971,Dover1972}. Note that the one-body Coulomb potential that plays an important role in the low-energy scattering is obtained self-consistently in the calculation. 

The partial scattering equations
\begin{align}
    \left\{\dfrac{\hbar^2}{2m'}\left[-\dfrac{d^2}{dr^2} + \dfrac{\ell(\ell+1)}{r^2} \right] + V(E,r) - E \right\} \nonumber \\
    \chi_\ell(E,r) = 0,
\end{align}
is solved with the Skyrme HF optical potential in Eq.~\eqref{Vopt} to obtain the scattering phase shifts and the excitation function using a computer program such as \texttt{ECIS06} \cite{ecis06}. 
The use of $m'=mA/(A-1)$ takes into account a large part of the center-of-mass correction important in the case of light nuclei. The $s$-wave scattering ($\ell = 0$) dominates at the energy of our interest. The width of the resonance $\Gamma(E_r)$ is calculated from the derivative of the phase shift $\delta_{\ell = 0}(E)$ with respect to energy at the position of the resonance $E_r$,
\begin{equation}
    \Gamma(E_r) = 2\left[\dfrac{d}{dE}\delta_{\ell = 0}(E)\right]^{-1}.
\end{equation}

\textit{Discussion.} Without any adjustment, the Skyrme HF calculation provides a good description of the background cross section and the resonance with the single-particle properties. The resonance recently found by Y. Ayyad \textit{et al.} \cite{Ayyad2022} is located at 182 keV that is close to the $s_{1/2}$ single-particle resonance with $J^\pi = 1/2^+$.
\begin{figure}[t]
    \centering
    \includegraphics[scale=0.5]{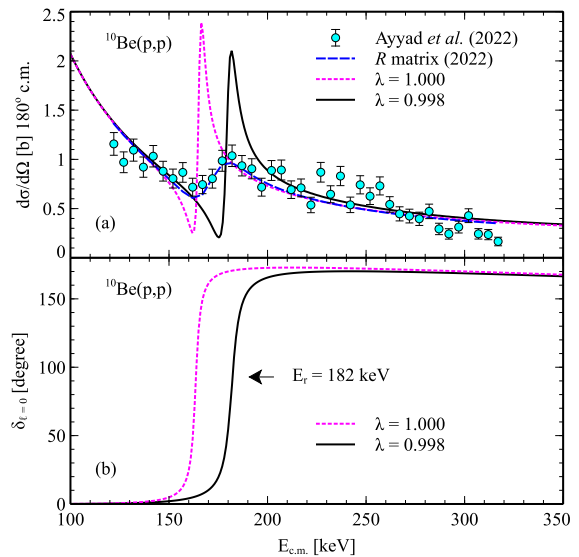}
    \caption{Calculated excitation function (a) and $s$-wave phase-shift (b) in the center-of-mass frame for the $^{10}$Be($p,p$) elastic scattering. The calculation with SAMi interaction is shown as an example. The experimental data and the $R$-matrix fitting (the blue-dashed line) are taken from Y. Ayyad \textit{et al.} \cite{Ayyad2022}.}
    \label{fig:CS}
\end{figure}

Our calculations with the SkM$^*$ and SAMi interactions show the resonance in the appropriate energy region. Outside the resonance, the calculation agrees well with experimental data. In general, the Skyrme HF calculation cannot predict exactly the location of the unknown resonance in $^{10}$Be($p,p$). Therefore, the nuclear central part of the Skyrme HF optical potential is slightly scaled by the factor $\lambda$ that is close to unity (Table \ref{tab1}) in order to correct the resonance position to 182 keV.
In the case of SLy4 and SGII interactions, the calculations with $\lambda = 1$ do not show the resonance in the energy region. As the resonance is just a few hundred keV above the threshold, it is possibly shifted to the sub-threshold energy with using SLy4 and SGII versions. However, with the slight correction (Table \ref{tab1}), the resonance at 182 keV appears for all selected variants of the Skyrme interactions.

\begin{table}[b]
\setlength{\tabcolsep}{10pt}
    \centering
    \caption{The scaling factor $\lambda$ in the calculations with different Skyrme interactions. The values of $\lambda$ are very close to unity. The calculated resonance width is about 5.5-6.0 keV.}
    \begin{tabular}{lcc}
    \hline\hline
        Skyrme interaction & $\lambda$ & $\Gamma$ [keV] \\
        \hline
        SkM$^*$ \cite{Bartel1982} & $1.004$ & $5.43$ \\
        SGII \cite{Giai1981} & $1.028$ & $5.97$ \\
        SLy4 \cite{Chabanat1998} & $0.984$ & $6.09$ \\
        SAMi \cite{Roca-Maza2012} & $0.998$ & $5.99$ \\
    \hline\hline
    \end{tabular}
    \label{tab1}
\end{table}

It is expected that the width of a near-threshold resonance is narrow. Indeed, the calculated resonance widths are 5-6 keV with different Skyrme interactions as seen in Table \ref{tab1}. Fig.~\ref{fig:CS} shows the calculation with and without the adjustment for the case of SAMi interaction as an example. Only the phase shift of the $s$-wave is shown in Fig.~\ref{fig:CS}(b) as the contributions of other waves are negligible. The resonance energy, $E_r = 166$ keV, and the width, $\Gamma = 4.05$ keV (the dotted line in Fig.~\ref{fig:CS}), are obtained with $\lambda = 1$. The calculations with different Skyrme interactions using the values of $\lambda$ in Table \ref{tab1} cannot be distinguished in Fig.~\ref{fig:CS}.

\begin{figure}[t]
    \centering
    \includegraphics[scale=0.5]{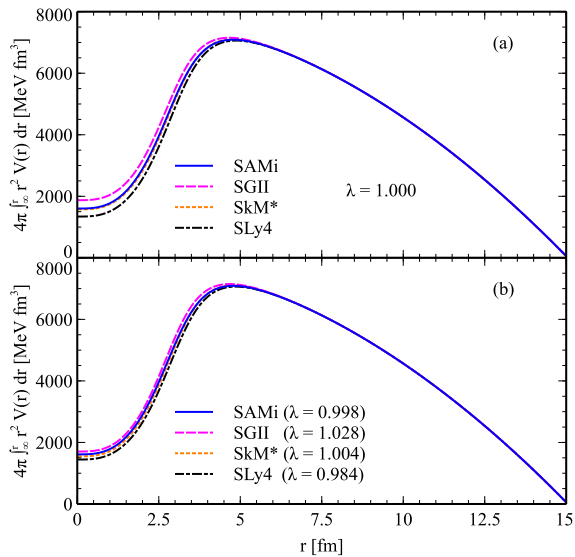}
    \caption{Illustration of the sensitivity of the resonance position to the accumulation of the volume integral of the total potential from the outer to the inner region. The calculations with different Skyrme interactions are convergent after the adjustment of the potential depth to correct the resonance position at 182 keV.}
    \label{fig:vol}
\end{figure}

As seen from Fig.~\ref{fig:CS} and Table \ref{tab1}, the position of the resonance is very sensitive to slight changes of the Skyrme HF potential. It can be understood by observing the accumulation of the volume integral. The value of this intergal, $4\pi\int_\infty^r V(r)r^2\,dr$, with $V(r)$ as the sum of the nuclear central and Coulomb potential, as a function of the relative distance $r$ is shown in Fig.~\ref{fig:vol}. As the potential only weakly depends on  energy, the illustration is done at zero energy. Note that the accumulation starts at the large distance where the total potential almost vanishes ($r = 15$ fm in this case). At the distances from 5 to 15 fm, only the effect of the Coulomb potential survives. At a distance where the nuclear potential plays a noticeable role (at about 5 fm, if $\lambda$ is unity), the difference in the calculations applying different Skyrme interactions can be observed clearly as shown in Fig.~\ref{fig:vol}(a). Therefore, the excitation functions for different Skyrme interactions vary. 

Figs.~\ref{fig:vol}(a) and (b) show that the small variation of $\lambda$ makes a large change in the accumulation of the volume integral. Consequently, the result of the calculation is changed. When the value of $\lambda$ is adjusted to correct the resonance location, all computations are getting very close. The calculations with different Skyrme interactions lead to the same properties of the studied $s_{1/2}$ single-particle resonance. Finally, we have to emphasize that slight changes of the HF potential caused by the scaling factor $\lambda$ do not affect the calculated single-particle structure and nuclear properties.

\textit{Conclusion.} The experimental result ~\cite{Ayyad2022} of the near-threshold resonance in $^{11}$B relevant to the $\beta$-delayed proton emission of $^{11}$Be is interpreted as the $s_{1/2}$ single-particle resonance predicted by the Skyrme HF calculation. Its simple structure is given by [$^{10}$Be($0_{\rm g.s.}^+$) + $p(s_{1/2})$].
The calculated width of this resonance is $\Gamma = 6$ keV and $J^\pi = 1/2^{+}$. The experiment in Ref.~\cite{Ayyad2022} is crucial for localizing the exact position of this resonance. The final result of calculations does not depend on the choice of the version of the Skyrme interactions. The Skyrme HF calculation is an excellent tool for describing the scattering process including low-lying single-particle resonances in the keV-energy region. 

The great current interest in experiments at rare isotope beam facilities around the world and in corresponding reaction theory is important for the progress in understanding unusual features of exotic nuclei in relation to nuclear structure, astrophysics, and the standard model of particle physics. Another branch of this physics is related to the role of clusters in physics of light and medium nuclei; this will be touched in the prepared FRIB experiment on the decay of $^{11}$Be into $^{7}$ Li and alpha-particle.

\textit{Acknowledgment.} B. M. L. was supported by the Institute for Basic Science (IBS-R031-D1).

\bibliography{refs}
\end{document}